# Expert System for Quality Assessment in "Tibiscus" University

Univ.Instr. Cătălin Țuican,
Assoc.Prof. Tiberiu Marius Karnyanszky, Ph.D.
Univ.Instr. Bogdan Ioan Șelariu
"Tibiscus" University of Timișoara, România

ABSTRACT. The periodical evaluation of the teaching staff in „Tibiscus" University bases on the specifications of The Romanian Agency for Quality Assurance in Higher Education (Agenția Română de Asigurare a Calității în Învățământul Superior, ARACIS) namely[1]:
*„The quality of the teaching and researching staff*
The universities must dispose of teaching staff which, as number and functional base, must be correctly allocated to the total number of students, depending on the study domain, and regarding the qualifications it must depend on the specific of the study program and the proposed quality objectives."
This paper presents the implementation of an expert system, offering to the students the possibility to perform the evaluation in a modern way and to the evaluation committee a quick access to all necessary data.

## 1. Introduction

Most Web applications have a client/server architecture, therefore to build a web application we need several things: a web browser (Internet Explorer, Mozilla Firefox), a web server responsible for the communication with the browser (Apache, IIS), a relational database server witch will store the information necessary for the application (MySQL, Oracle), and a database governing system (SGBD) and a programming language to link the web server and the database server. For all this to work we need an operating system compatible wit the web server, the programming language and the database server.

---

[1] [Ara06] p. 37

The aim to implement such an application is based on the activity of periodical evaluation of the teaching staff. According to the ARACIS principles, the evaluation procedure elaborated by the Commission of Quality Assessment and Assurance in the "Tibiscus" University (Comisia de Evaluare și Asigurare a Calității în Universitatea „Tibiscus", CEACUT) and applied by the faculty Quality Commission (Comisia Calității, CC), count on the following regulations[2]:

*„The students' evaluation of the teaching staff*

Minimum standard: A questionnaire for the students' evaluation of the teaching staff is available, it is approved by the university senate, it's applied on demand at the end of each semester and the application's results are confidential and accessible only to the dean, the rector and the evaluated teacher.

Optimal standard: The evaluation is compulsory. The results of the student's evaluation are separately discussed, statistically processed apart on chairs, faculties and university, to obtain a transparent policy on the educational quality."

## 2. Describing the application

We've named the application "Online Evaluation" witch outlines the utility of multimedia databases and for it's creation we used the MySQL database server, the PHP programming language and the Apache server. The application offers the student a web interface that permits communication between it and a database to elaborate statistics concerning the reactions of the student to the way professors from higher educational institutions go about their activities.

      The program lists the utterances of the evaluation questionnaire one after the other and waits for the student's response. The application witch we are presenting in this project is an illustration of an interactive program

---

[2] [Ara06] p. 38

category capable of taking some information from a database and then processing that information. In principle the connections to a database (in our case MySQL) can be realized simply using a few PHP modules.

The .html pages were created using the Adobe Dreamwaver utility, and the ones containing php codes were saved as .php. All operations – insertions, modifications, deleting data – that are done to the database are created using php. To have a pleasant and dynamic page we used .jpg images and Flash video clips, created with Adobe Photoshop and Adobe Premiere.

It's true that there are sites that have evaluation programs integrated within them but this application is conceived in the hope that it will be used in our University sometime in the future.

There are two types of users which can access this program. One of the users is the administrator, this can be one person or several professors which can access and implement information from/to the database, and the other type of user (the student) which besides the fact that he can do an evaluation he can also see how many evaluations have been done as well as the result of the evaluations done by other students.

Running on the localhost server, maintenance of the database as well as manipulation of it are very easy because it can be done using phpMyAdmin running on localhost as root@localhost.

The administrator can introduce new data, modify the state of the application, to erase or put in new teachers, to modify all sorts of records, in essence he has absolute power over the database.

The application consists of defining a database necessary to store information about the administrator, professors, questions, IP's, etc. and creating a user interface. The most important thing for an application created for users is it's interface with them. The user interface must represent in a very explicit manner the functions it is interfacing.

First of all we need to install the software: PHP, MySQL and Apache, and the MySQL and Apache servers must be started. We also need a text editor, Word or Notepad are good examples.

The next important step is creating the database to store the data after which using the PHP programming language we will prevail the data from the database and display them accordingly.

### 2.1. Connecting to the database

Every time we manipulate data from a database using php code, we have to specify which database we are using, in other words we have to connect to the database that contains the information we want to manipulate.

Because we used more files that collect information from the database we used for the connection a connect.php script through which we will make the connection to the database and which we will include where it is needed.

The lines of the connect.php file are:

```php
<?php

  $con = mysql_connect("localhost","root","");

  $nume_bd = mysql_select_db('evaluare')

?>
```

where: mysql_connect("localhost", "id", "password"); ensures the connection to the Web server, being identified by an ID and a password;

mysql_select_db('evaluare'); ensures selection of the "evaluare" database only if the connection to the Web server was successful.

The files which will be included can contain everything which a normal php script would contain, HTML tags, PHP functions, PHP classes. If these instructions show up multiple times in a script, the file will be included each time.

### 2.2. Designing the database

In building a database we need to follow a process of identification and organization of columns and grouping the columns into tables (entities). The database for the interactive application "Online Evaluation" is comprised of six tables.

We made this database using phpMyAdmin.

The "rezultate" table – represents the final results of the evaluation

The "eval_sesiune" table – represents the intermediary results of an evaluation session

The "listaprofi" table – contains the information about the professors

The "listaip" table – contains the list of IP's on which the evaluation can take place, in the present case the only IP introduced is 127.0.0.1 representing the localhost

The "admin" table – this is where we store the information about the administrator

The "stare" table – this is where the information about the evaluation session are stored.

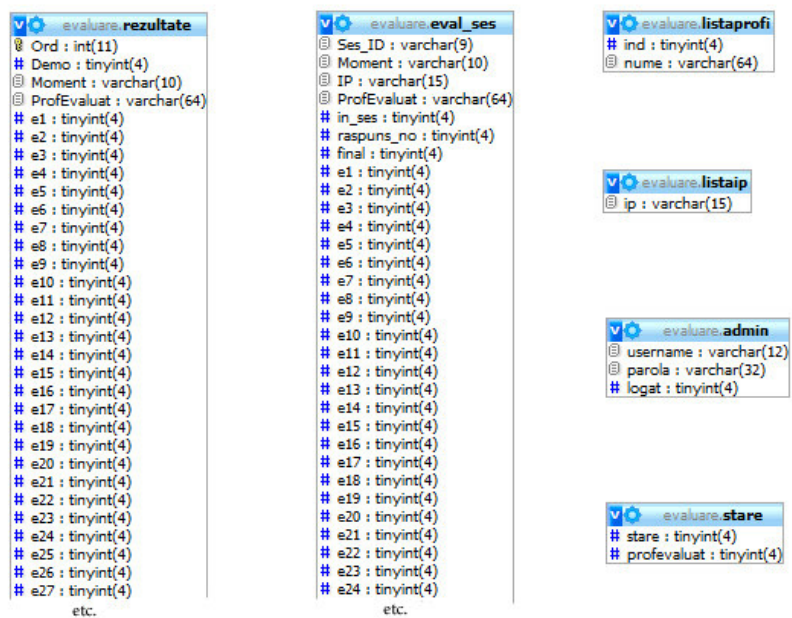

*Figure 1. The tables of the evaluation database*

## 2.3. The application functionality

### 2.3.1 Functioning at administrator level

We need an administrative section whit the help of which we can add or modify access rights for the users of the database, as well as the access to adding/modifying/deleting data about the teachers.

243

The administrator page can only be accessed via a username and password.

After connecting the administrator will have access to a menu from which he can choose whatever option he wants.

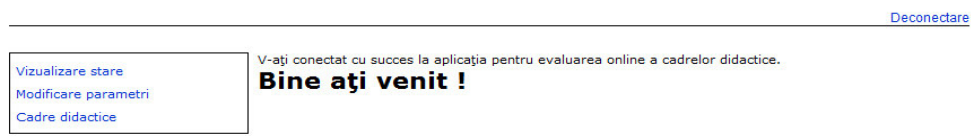

*Figure 2. The interface with the administrator – options*

*The View state menu* – By using this menu we can view characteristics about the state of the application such as the state of the evaluation active/inactive, the professor being evaluated or the list of preferred IP's.

*The Parameter Modification menu* – By entering this menu you can edit/modify the characteristics of the state of the application. The state of the program can be selected to be inactive (the test version) or active. If it is selected to be active then we can also select the professor evaluated from a list in the database. We can also add or modify the list of IP addresses from which the evaluation can be made when the program's state is active. For all the modifications to take effect at the end we most press on the "Actualizati" button.

*The Professors menu* – Using this menu we can add, modify or delete a professor. To add a new professor we introduce the full name, and then we upload from the computer a photo of the professor in .jpg format after which we press the button "Adaugati". From the Professors menu we can enter the "modificati" submenu located on the right hand side under each professor's photo. Once we access this submenu we can change the professors name and the photo after which we press "Actualizati". We can also delete the whole profile of the teacher by checking the „Ștergeți acest profil" checkbox after which we press the "Actualizati" button.

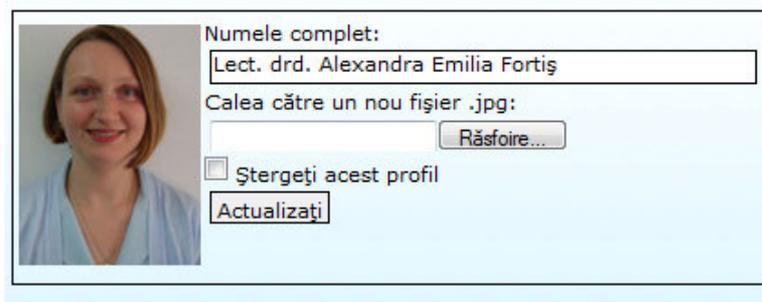

*Figure 3. Modifying teachers*

### 2.3.2. Functioning at user level

The user can access the application via a web address decided by the administrator. Once the application accessed, the user is informed about the minimal requirements, a short presentation of the program and the functioning of the program. There is also a warning message about the fact that once an answer given the user can no longer return to change the answer if he has proceeded to a different question.

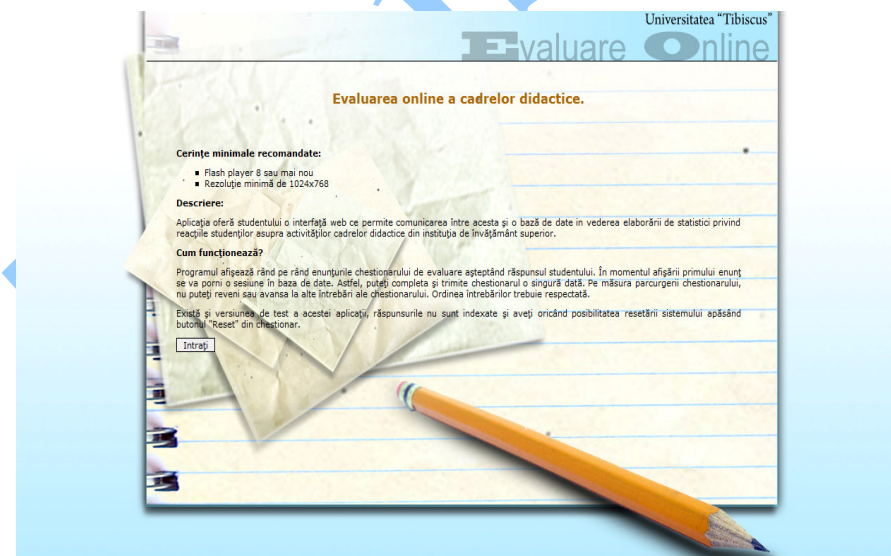

*Figure 4. The user page*

After pressing the button "Intrati" the user gets access to an introductory page of the program. This is where you find information on how to fill out the evaluation.

This questionnaire contains 58 questions which describe behavior habits of the professors which teach at the university. Read each question carefully and decide in which way it applies to the professor being evaluated and please answer whit one of the following variants: Very Poor, Poor, Medium, Good, and Very Good.

On the form which is being shown at any one time, select the radio button on the same level as the answer you want to give. Answer all the questions! Try to be objective! Tri to answer each question without being affected by the general image you have about the professor you are evaluating. This project is implemented in such a manner as to ensure the anonymity of the student. The evaluations you've made will be processed statistically together whit the evaluations done by the other students and only the results will be shown. To begin completing the questionnaire press "Continuati".

The program shows one after the other the questions waiting for the student answer. At the moment the first question is shown a session is started in the database. In this way you can fill out and send the questionnaire only once. Once the questionnaire started you cannot return to or fast forward to other questions.

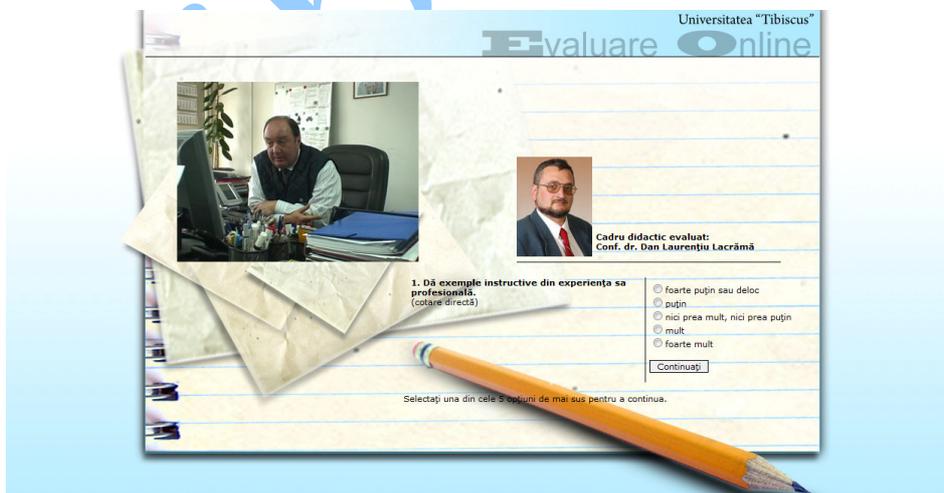

*Figure 5. Questions*

After filling out all the questions, an end message shows up on the screen and a link to the page which allows you to view your answers.

The user has the possibility to view the results of the questionnaire. To view a questionnaire separately or to print it the user has to click the symbol ">>" at the same level as the record he/she is interested in.

*Figure 6. Example of a questionnaire*

**Conclusions**

This paper presents the teacher's evaluation system applied at the "Tibiscus" University of Timişoara, based on the legal demands of the Romanian Agency for Quality Assurance in Higher Education. This evaluation offers a mark (Very Good, Good, Medium, Poor, Very Poor) for the scientific competence, the psycho pedagogical competence, the psychosocial competence and the managerial competence. A database memorizes the students' answers and a computer application analyzes and processes all valid data. These results of the student's evaluation are further discussed and statistically processed, apart on chair, faculty and university. All results are presented to the faculty management and the evaluated person, then worked out to obtain the quality evaluation of the studying program, document presented into the faculty/university and published to ensure a transparent policy of the educational quality.

References

[Ara06]   Agenția Română de Evaluarea Calității în Învățământul Superior – *Metodologia de evaluare externă, standardele, standardele de referință și lista indicatorilor de performanță a Agenției Române de Asigurare a Calității în Învățământul Superior*, București, 2006.

[Dra08]   N. Drăgulănescu – Motivații și obstacole ale asigurării calității în învățământul superior, on http://mepopa.com

[Pan08]   Ioan Pânzaru – *Asigurarea calității în învățământul superior din țările Uniunii Europene*, București, 2008, on http://www.romaniaeuropa.com/cartionline/carti_pedagogie/asigurarea_calitatii_in_invatamantul_superior_din_tarile_uniunii_europ.php

[Pav07]   M. Pavelescu – *Calitatea învățământului. Limitele standardizării,* București, 2007, pe http://www.poezie.ro/index.php/press/224147/index.html

[***06]   *** – *Hotărârea de Guvern nr. 1418/2006 pentru aprobarea Metodologiei de evaluare externă, a standardelor, a standardelor de referință și a listei indicatorilor de performanță a Agenției Române de Asigurare a Calității în Învățământul Superior*